\documentstyle[epsf,11pt]{article}

\begin{document}

\vspace*{-1.8cm}
\begin{flushright}
{\large\bf LAL/RT 06-12}\\
\vspace*{0.1cm}
{\large September 2006}
\end{flushright}
\vspace*{1cm}

\begin{center}
{\LARGE\bf New final doublets and power densities for the ILC\\
\vspace*{0,2cm}
 small crossing angle layout}
\end{center}
\vspace*{1cm}

\begin{center}
{\large\bf R. Appleby$^a$\footnote{Talk given at the Linear Collider Workshop ``{\it\bf LCWS06}'',9-13 March 2006, I.I.Sc Bangalore, India}}\\
\vspace*{0,2cm}
$^a$ {\bf The University of Manchester and the Cockcroft Institute}\\
Oxford Road, Manchester, M13 9PL, UK\\
\vspace*{1cm}
{\large\bf P. Bambade$^b$}\\
\vspace*{0,2cm}
$^b$ {\bf Laboratoire de l'Acc\'{e}l\'{e}rateur Lin\'{e}aire}\\ 
CNRS/IN2P3 and Universit\'{e} Paris Sud 11, B\^{a}t. 200, B.P. 34, F-91898 Orsay, Cedex, France
\end{center}
\vspace*{0,8cm}

\begin{abstract}
In this paper we use current and proposed final doublet magnet technologies
to reoptimise the interaction region of the International Linear Collider and reduce the power losses. The result is a set of
three new final doublet layouts with improved beam transport properties. The effect of localised
power deposition is considered, and its reduction using Tungsten liners.
\end{abstract}
\vspace*{0,8cm}

\section{Introduction}

The current baseline configuration of the International Linear Collider consists of two interaction regions (IRs) and two detectors, with one IR having a large beam crossing angle and
one having a small beam crossing angle. In this paper we are concerned with the small crossing angle layout, which presents considerable technical challenge, mainly
resulting from transporting the outgoing disrupted beam off-axis in the final doublet. The first presentation of the
downstream extraction line optics was made at Snowmass 2005~\cite{2mradsnowmass}. This layout was developed
for 1 TeV using NbTi superconducting final doublet magnets and assumed that for 500 GeV all fields
would be scaled down.

The current (baseline) 2mrad design shows unacceptable beam power loss, both in the
final doublet and in the extraction line, for some of the beam parameter sets considered for the ILC. These parameter sets were documented in ~\cite{schulte} 
and are designed to be a set of representative beam parameters to assure 
flexibility in case of unexpected practical limitations in achieving some of the machine goals. 

In this paper, we exploit the limits of current magnet technology, as well as technology still under development,
to optimise the 2mrad final doublet layout. We present three new final doublet designs, one optimised for
the baseline energy of 500 GeV and two for the upgrade energy of 1 TeV, using as criterion the combined power deposition from charged beam and radiative Bhabha particle 
losses and the worst case among the different beam parameters. 

\section{The baseline layout and the magnet technologies}

The final doublet in the 2mrad scheme is based on a superconductive large bore QD0 magnet and on a normal conducting smaller aperture QF1 magnet. The two nearby 
sextupoles for local chromaticity correction, SD0 and SF1, are also superconductive large bore magnets. In the current design, the choice of superconductive technology for 
QD0 is NbTi and the maximum pole tip field which is assumed to be achievable is 5.6 T. 

In this work, we propose to use NbTi magnets~\cite{saclay} with a somewhat larger
maximum pole tip field of 6.28 T (accounting also for expected
field-reducing effects from the detector solenoid and for a safety margin)
and optimise the design separately for the 500 GeV and 1 TeV machines. We
also investigate using Nb3Sn-based technology for QD0 at 1 TeV. In this
case, the maximum pole tip field assumed is 8.8 Ts~\cite{saclay}. In all cases, the
sextupoles are also required to be large bore superconductive magnets. At
present, NbTi technology is assumed for these with similar features as in
the baseline design, but partially improved parameters. In all three
designs, the shortening of the magnets which these larger pole tip fields
allow result in much improved beam transport properties as compared with
the present design, for both the 500 GeV and 1 TeV designs. The
optimisation procedure is described in more detail in~\cite{eurotevfd}.

\section{The results of the final doublet reoptimisation}

The results of the optimisation can be found in table~\ref{tabmpnbti500} for NbTi at 500 GeV and in table~\ref{tabmpnb3sn1000} for Nb$_3$Sn at 1 TeV.
The detailed results for the power losses and the magnet parameters for NbTi at 1 TeV can be found in\cite{eurotevfd}.

\begin{table}[h]
\vspace{-4mm}
\begin{center}
\caption{\it Quadrupole and sextupole parameters at $\sqrt{s}=$500 GeV for NbTi. In this table 
the sextupole apertures have been reduced to the minimum required for zero-loss transport of the
disrupted beam, assuming the worst case among the beam parameter sets,
including with transverse offsets maximising the beamstrahlung radiation.}
\vspace{5mm}
\begin{tabular}{|l|c|c|c|c|c|}
\hline
Magnet	& Length  & Strength & radial aperture & gradient & B$^{\mathrm{PT}}$  \\ \hline \hline
QD0 	& 1.23m & -0.194 m$^{-1}$    & 39mm		  & 161.6 T m$^{\mathrm{-1}}$ & 6.30 T    	\\ \hline	
SD0	& 2.5m	& 1.117 m$^{-2}$    & 76mm		  & -	     & 2.69 T 	   \\ \hline
QF1	& 1.0m	& 0.082 m$^{-1}$    & 15mm		  & 67.9 T m$^{\mathrm{-1}}$   & 1.02 T   \\ \hline
SF1	& 2.5m	& -0.273 m$^{-2}$   & 151mm		  & -	     & 2.59T   	   	\\ \hline
\end{tabular}
\label{tabmpnbti500}
\end{center}
\end{table}
\vspace*{-0,9cm}
\begin{table}[h]
\vspace{-0mm}
\begin{center}
\caption{\it Quadrupole and sextupole parameters at $\sqrt{s}=$1 TeV for Nb$_3$Sn. In this table 
the sextupole apertures have been reduced to the minimum required for zero-loss transport of the
disrupted beam, assuming the worst case among the beam parameter sets,
including with transverse offsets maximising the beamstrahlung radiation. Note the pole-tip field limit of the first sextupole is exceeded.}
\vspace{5mm}
\begin{tabular}{|l|c|c|c|c|c|}
\hline
Magnet	& Length  & Strength & radial aperture & gradient & B$^{\mathrm{PT}}$  \\ \hline \hline
QD0 	& 2.0m & -0.120 m$^{-1}$   & 44mm		  & 200 T m$^{\mathrm{-1}}$  & 8.80 T 	\\ \hline	
SD0	& 3.8m	& 0.646 m$^{-2}$    & 95.1mm  		  & -	     & 4.87 T         \\ \hline
QF1	& 2.0m	& 0.041 m$^{-1}$    & 15mm		  & 67.8 T m$^{\mathrm{-1}}$   & 1.02 T \\ \hline
SF1	& 3.8m	& -0.169 m$^{-2}$   & 163mm		  & -	     &	3.74 T    	   	\\ \hline
\end{tabular}
\label{tabmpnb3sn1000}
\end{center}
\end{table}

It was shown that in the case of the 500 GeV machine, all
ILC beam parameter sets can be accommodated with the new design, as long as
3 mm thick Tungsten liners~\cite{eurotevfd} are included in the design of the vacuum chamber. These liners spread out
the showers and reduce the maximum power density at the location where most
losses are concentrated near the outer edge of QD0 to values lower than
the specified tolerance of 0.5 mW/g. In the case of the 1
TeV machine, on the other hand, it was found that while for all cases, the
new final doublet layouts significantly improve the beam transport
properties, a further improvement in gradient, of about 20\% beyond that
assumed with Nb$_3$Sn, would be needed to remain within the same tolerances to
avoid magnet quenching.

\section{Conclusion}

This work shows that the layout of the 2mrad design can be optimised at 500 GeV with NbTi technology to obtain
small enough beam power depositions in the final doublet.
This doublet now needs to be integrated into the final focus system and extraction line. 
When this will be done, we will propose that it become the baseline for the 2mrad at 500GeV, with the assumption that it would have to 
be replaced at the time of the 1 TeV energy upgrade. In all cases, the sextupole lengths, strengths and apertures are
not fully optimised. Some more work is still needed here, in view of reducing their sizes and ease their integration with the detector.

A final doublet based on NbTi technology does not, on the other hand, provide small enough beam power depositions at 1 TeV, for several of the beam parameter sets considered. 
Exploiting Nb$_3$Sn superconductive technology can however improve the situation considerably.
The optimised layout shows far superior power loss behavior than the current
1 TeV machine baseline final doublet, but is still in excess of the quenching limit for the beam parameter sets with the 
largest beamstrahlung energy loss. It was estimated that about a
factor 1.2 larger gradient than that assumed with Nb3Sn would be needed to
satisfy the corresponding tolerance in the QD0 magnet. We expect further R\&D in superconductive magnet technology to enable such increases in maximum available pole tip field in the future, which will ease the design of the 2mrad beam crossing-angle layout. An alternative path is to constrain the optimisation of the beam parameters at 1 TeV with an upper bound on the beamstrahlung energy loss. This would also be motivated since a very large beamstrahlung energy loss is undesirable for several other reasons.

\end{document}